Doubting, Testing, and Confirming Galileo:

A translation of Giovanni Battista Riccioli's experiments regarding the motion of a falling body, as reported in his 1651 *Almagestum Novum*

Christopher M. Graney

Jefferson Community & Technical College

Louisville, Kentucky (USA), 40272

christopher.graney@kctcs.edu

The Italian astronomer Giovanni Battista Riccioli is commonly credited with performing the first precise experiments to determine the acceleration of a freely falling body.  Riccioli has been discussed by historians of science, sometimes positively but often not, but translations of his work into modern languages are not readily available.  Presented here is a translation of his experiments regarding the nature of the motion of a falling body.  Riccioli provides a thorough description of his experiments, and his data are quite good.  He appears to have a model approach to science: He attacks the question of free fall with the expectation of disproving Galileo's ideas, yet he is convinced by his data that Galileo is indeed correct, and he promptly informs a former protégée of Galileo's of the results.

Giovanni Battista Riccioli (1598-1671), an Italian astronomer, published his encyclopedic treatise *Almagestum Novum* in 1651. This was an influential, massive work whose length exceeded 1500 large-format pages, mostly of dense type with some diagrams. Within it, Riccioli discusses a wide variety of subjects; one of these is the behavior of falling bodies, including the results of extensive experiments.

These experiments are often cited as the first precise experiments to determine the acceleration due to gravity [Koyré 1953, 231-232; Koyré 1955, 349; Lindberg & Numbers 1986, 155; Heilbron 1999, 180]. In his falling body experiments, Riccioli reported limited agreement with Galileo, which was significant because Riccioli was a prominent figure who opposed Galileo in many ways, going so far as to include the text of Galileo's condemnation in his work (Meli 2006, 134). These experiments have been discussed elsewhere in some detail (Koyré 1953; Koyré 1955; Heilbron 1999) but full translations of Riccioli's work from Latin into modern languages are not readily available.

Riccioli was an Italian Jesuit working in Bologna, Italy with other scientifically-inclined Jesuits, most notably Francesco Maria Grimaldi (1618-1663). Riccioli was a geocentrist, but Edward Grant has noted that, unlike other geocentrists who

> ...were not scientists properly speaking but natural philosophers in the medieval sense using problems in Aristotle's *De caelo* and *Physics* as the vehicle for their discussions, Riccioli was a technical astronomer and scientist.... [Grant 1984, 12]

The *Almagestum Novum* reflects this. It is filled with extensive reports on experiments and tables of data from real experiments, reported whether the data fit a particular model or not. It reflects close, careful work, including the work with falling bodies conducted so as to determine their behavior experimentally and the work necessary to determine that only small-amplitude oscillations of a pendulum are isochronous while larger oscillations have longer periods. Riccioli often illustrates the reliability of his work by providing descriptions of how it was carried out. Thus those who wished to reproduce the experiments in the *Almagestum Novum* could do so (Meli 2006, 131-134). Riccioli is probably best known for the maps of the Moon included in the *Almagestum Novum*. These maps, produced by Grimaldi and Riccioli, introduced the system of lunar nomenclature still used today. Indeed, the "Sea of Tranquility" ("Mare Tranquillitatis"), which became an icon of modern culture in 1969 when the Apollo 11 "Eagle" landed there, was named by Riccioli (Bolt 2007, 60). The Moon maps again reflect thoroughness and attention to detail. Alexandre Koyré provides a fine illustration of Riccioli's almost obsessive concern for detail and accuracy in his works in the following discussion of Riccioli's efforts to calibrate a pendulum whose strokes would measure out precise seconds, and which could serve as a standard against which quicker pendulums could be calibrated:

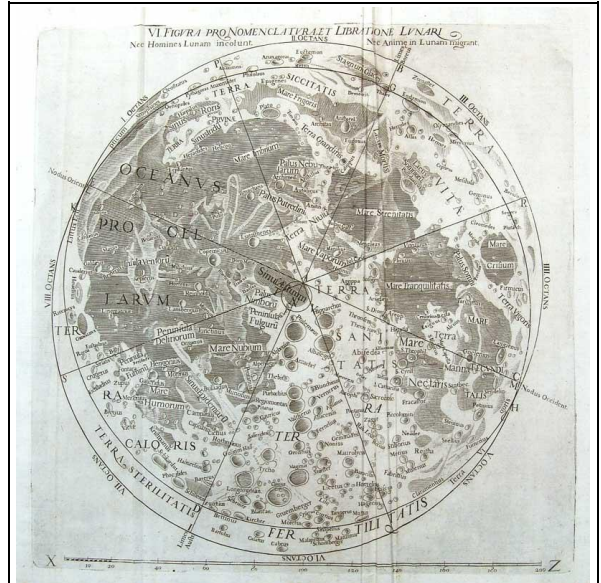

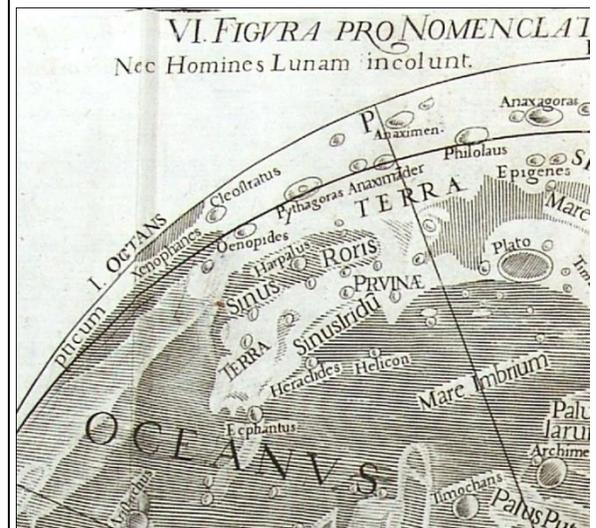

**Top:** Moon map from Riccioli's *Almagestum Novum*.
**Bottom:** Map detail.

> For six consecutive hours, from nine o'clock in the morning to three o'clock in the afternoon, he counts (he is aided by the R. P. Francesco Maria Grimaldi) the oscillations. The result is disastrous: 21,706 oscillations instead of 21,660. Moreover, Riccioli recognizes that for his aim the sundial itself lacks the wanted precision. Another pendulum is prepared and "with the aid of nine Jesuit fathers," he starts counting anew; this time — the second of April 1642 — for twenty-four consecutive hours, from noon to noon: the result is 87,998 oscillations whereas the solar day contains only 86,640 seconds.

Riccioli makes then a third pendulum, lengthening the suspension chain to 3 feet, 4.2 inches. And, in order to increase the precision even more, he decides to take as a unit of time not the solar, but the sidereal day. The count goes on from the passage through the meridian line of the tail of the Lion (the twelfth of May 1642) till its next passage on the thirteenth. Once more a failure: 86,999 oscillations instead of 86,400 that there should have been.

Disappointed yet still unbeaten, Riccioli decides to make a fourth trial, with a fourth pendulum, somewhat shorter this time, of 3 feet, and 2.67 inches only. But he cannot impose upon his nine companions the dreary and wearisome task of counting the swings.  Father Zeno and Father F. M. Grimaldi alone remain faithful to him to the end.  Three times, three nights, the nineteenth and the twenty-eighth of May and the second of June 1645, they count the vibrations from the passage through the meridian line of the Spica (of Virgo) to that of Arcturus. The numbers are twice 3,212 and the third time 3,214 for 3,192 seconds.  [Koyré 1953, 230]

Yet history has not been generous to Riccioli.  For example, Riccioli devoted a large portion of the *Almagestum Novum* to the question of the mobility or immobility of the Earth, which Grant calls

...the lengthiest, most penetrating, and authoritative analysis made by any author of the sixteenth and seventeenth centuries. [Grant 1996, 652]

Nonetheless, Riccioli's analysis has often been dismissed as a matter of deciding the question by weighing arguments by quantity rather than by quality, or by invoking religious authority rather than good science, even though reading the *Almagestum Novum* reveals the contrary to be true (Graney 2012; Graney 2011; Graney 2010).  As another example, Riccioli has often been portrayed as a "secret" scientist who for religious reasons hid his true thinking (Dinis 2002).

Thus a direct translation of Riccioli's falling body experiments, rather than additional discussion about them, is of value.  The text translated here is headings II and III in Chapter XXVI of Book IX, Section IV of the second volume of the *Almagestum Novum*, pages 384-7; these headings and pages contain the experiments to determine the nature of motion of a falling body. Included as side-notes with the translation is some brief commentary on Riccioli's work, as well as a basic modern scientific analysis of his data.

We (I thank Christina Graney for her invaluable assistance in translating Riccioli's Latin) intend this translation to hew closely to Riccioli's original work.  However, Riccioli uses many different sorts of verbs (pluperfect active subjunctive!), sentences that run in excess of 100 words in length, and few paragraphs, among other things.  Therefore, we use simpler verbs.  We turn long sentences into multiple short sentences.  We break long paragraphs into short ones (a double-space indicates Riccioli's original paragraphs; indents indicate where we add a paragraph break).  We retain Riccioli's numbering system, which uses one set of Roman numerals for paragraphs and another for headings.

*II. The Group of Experiments About Unequal motion of Heavy Bodies descending faster and faster in the Air, by which they come nearer to the end to which they tend.*

VI. The *first* Experiment is taken from sound.  Let a ball of wood or bone or metal fall from a height of 10 feet into an underlying bowl, and attend to the ringing arising from the percussion.  Then let that same ball fall from a height of 20 feet, and indeed you will perceive a far greater and more extensive sound poured out.  Afterwards lift up the bowl to a height of 10 feet and into that let drop the same ball from an altitude of 10 feet above the bowl.  Indeed you will perceive a ringing like at first.  Therefore that ball in the second fall has acquired a greater impetus[1] because of the drop from the greater height, than from the smaller heights in the first and third falls.  And in the second fall the ball has gained more impetus in the second half of the journey down than in the first half, by as much as its downward velocity will have increased in the latter 10 feet of the fall, than in the former 10 feet.

In fact it is manifest by continual experiments that in a moving body greater impetus accompanies swifter motion.  A household experiment with this phenomenon is to pour out water into a ladle from a vessel about two or three fingers' breadth from the ladle.  You will perceive no noise.  Elevate the vessel to two or three feet, and you will perceive noise from the falling

> **Riccioli is writing in a time when the most basic behaviors of falling bodies could still be disputed.  Thus he begins by establishing basic behaviors.  Here he illustrates simple experiments to show that the impetus-momentum gained by a falling body increases with the distance traversed in the fall, and that it is not dependent in an absolute way on the point of release.**

---

[1]  The concept of impetus, as discussed by Jean Buridan (1300-1358), is similar to the modern idea of momentum in that it is a product of mass and velocity and is directional.  However, Buridan did not distinguish between translational and angular momentum, viewing rotation as merely another direction (Clagett 1974, 275-7).

water.  Hence Cicero in *Somnio Scipionis* reports that people near the cataracts of the Nile have been deafened because of the crashing of the water falling from the precipitous height.

The *second* Experiment it is taken from the impact perceptible by the sense of touch.  Place your hand below a ball while someone lets it fall from an altitude of 10 feet.  Indeed you will experience the lightest impact.  But if the same ball is let fall from an altitude of 50 feet or greater, your hand will perceive some pain from the impact: therefore a greater impetus is derived from a higher fall.  Poor Aeschylus[2] sensed this greater impetus that I have described, from the turtle that the Eagle dropped onto his head; and the stupid bird herself felt no doubt concerning what would happen.  Elpenor[3] sensed it in falling from the tower. So Ovid writes in the third book of *Tristia*

> *Who falls on level ground – though this scarce happens –*
> *so falls that he can rise from the ground he has touched,*
> *but poor Elpenor who fell from the high roof*
> *met his King a crippled shade.[4]*

And from this source we know that adage and well-known warning of the Poet

> *...The higher they are raised,*
> *The harder they will fall...*

Finally, is it not true that they who run down a slope receive so great an impetus that, however much they may wish, they cannot

---

[2] Aeschylus:  An ancient Greek tragedian who, according to legend, was killed when an eagle mistook his bald head for a stone and dropped a tortoise on it in an attempt to break the tortoise's shell.

[3] One of Odysseus's crew.

[4] For the Ovid translation we have used Wheeler 1939, 117.

stop their forward movement at the bottom, even though they could easily stop it at the beginning?

VII. The *Third* Experiment is taken from the greater impact of falling from a height, but estimable by the eyes:  Namely a clay ball released from a small altitude is not itself broken, neither can it break the shell of an egg or the hull of a nut placed perpendicularly under it, nor can it elevate a weight placed in a wooden two-pan balance [when it strikes the other pan], nor penetrate a palm's-depth of water; if it is released from a higher place, it does all those things: namely, it breaks, and it is broken, and it elevates that weight.  Thus a wooden ball, or playing ball, falling from low altitude into a cistern or a large vessel full of water, is immersed a few finger's breadth underneath the water; but if that ball may plummet from a very high place, it penetrates to many feet below the water and finally all the way to the bottom. And by other innumerable experiments of this sort it is made evident that a heavy body falling from a higher place always naturally accrues greater and greater impetus at the end of its motion.

The *Fourth* is taken from bouncing, and by the rebound in the height of a playing ball.  Indeed we arranged for a very hard leather ball of this sort, no greater in size than the yolk of an egg, to be released in order that it might fall to the earth at an acute angle from an altitude of 37 feet, down to the flat stone pavement. It rebounded all the way to 7 ½ feet.  When it was released from an altitude of 73 feet it rebounded to 11 ¼ feet.  A larger leather ball released from an altitude of 37 feet and hitting the pavement by a more obtuse angle rebounded to nearly 6 feet; released from an altitude of 73 feet it rebounded to 7 ½ feet.

But I see myself as playing, as long as I am not progressing towards more noble experiments, and towards clearly demonstrating not only the non-constant motion of heavy bodies, but also the true growth of their velocity, which increases by uniform differences as the motion progresses.

VIII. Then the *Fifth* Experiment often taken up by us, has been the measuring of the space which any heavy body traverses in natural descent [i.e. free fall] during equal time intervals.  This was tested at Ferrara by Fr. [Niccolò] Cabeo in 1634, but only from the tower of our church there (less than 100 feet in height) and using an uncalibrated Pendulum.

But in 1640 in Bologna I calibrated Pendulums of various lengths using the transit of Fixed Stars through the middle of the heavens.  For this [free fall] experiment I have selected the smallest one, whose length measured to the center of the little bob is one and fifteen hundredths of the twelfth part of an old Roman foot, and a single stroke of which [that is, the half period] equals one sixth of a second, as I have shown and set forth in book 2 chapter 21.  As a single Second exactly equals six such strokes, then one single stroke is nearly equal to that time with respect to which the notes of semichromatic music are usually marked, if the Choirmaster directs the voices by the usual measure.

Riccioli provides a detailed description of the procedure used in his falling body experiments.

The oscillations or strokes of so short a Pendulum are very fast and frequent, and yet I would accept neither a single counting error nor any confusion or fallacious numbering on account of the eye. Thus our customary method was to count from one to ten using the concise words of the common Italian of Bologna (*Vn, du, tri, quatr, cinq, sei, sett, ott, nov, dies*), repeating the count from one, and noting each decade of pendulum strokes by raising fingers from a clenched hand. If you set this to semichromatic music as I discussed above, and follow the regular musical beat, you will mark time as nearly as possible to the time marked by a single stroke of our Pendulum. We had trained others in this method, especially Frs. Francesco Maria Grimaldi and Giorgio Cassiani, whom I have greatly employed in the experiment I shall now explain.

Grimaldi, Cassiani, and I used two Pendulums; Grimaldi and Cassiani stood together in the summit of the Asinelli Tower [in Bologna], I on the pavement of the underlying base or parapet of the tower; each noted separately on a leaf of paper the number of pendulum strokes that passed while a heavy body was descending from the summit to the pavement. In repeated experiments, the difference between us never reached one whole little stroke. I know that few will find that credible, yet truly I testify it to have been thus, and the aforementioned Jesuit Fathers will attest to this. That is all concerning the Pendulum and the measure of time.

> Here Riccioli provides a simple estimate of experimental error.

After this we prepared a very great basket full of clay balls, each of which weighed eight ounces. For the shorter distance measures at least, we used the windows of our College. But for the higher ones, we used windows or openings of different towers: the tower of St. Francis, which is 150 Roman feet high; and St. James, which is 189 feet; and St. Petronius of Bologna, which is 200 feet; and St. Peter, which is 208 feet; but especially and more frequently the tower of Asinelli, which is 312 feet high altogether, and 280 feet from the summit to the base or parapet. The Asinelli is as commodious as possible to this sort of experiment, just as if it might have been constructed for this purpose. It is a delight to the eye.

IX. Shown in the diagram below is the rather thick trunk of the Tower IBCD over an almost cubic base VYZX that is much broader than the trunk of the tower. From this base the parapet YZH stands out, fenced in by wide stone railings. At least six men may safely walk abreast around the tower, between it and the railings. On top, the crown BC stands out, fenced in by peaked stone railings. Thus any man of ordinary stature may be able to look over safely from the railings at G, Φ, and O, as well as from the windows. From there a plumbline can be dropped all the way to the pavement of the parapet ID to measure the height GI. Fr. Grimaldi and I have done this more than once, obtaining the value I stated earlier of 280 old Roman feet. We measured the rest as well.

Thus this sort of Tower has great advantages for this sort of work. For instance, balls released from openings G, Φ, and O fall perpendicularly to the pavement ID, neither impinging on the foot of the tower, nor falling outside the railings YZ. The balls do not fall out into the street by the base of the tower. Balls can be released from the crown often and frequently without danger to anyone. In addition, the tower has iron belts around F and T, with iron clasps, constraining opposite walls.  We have used these as reference points for measurements as well.  The line NH indicates heights which we have used, including those at other towers.

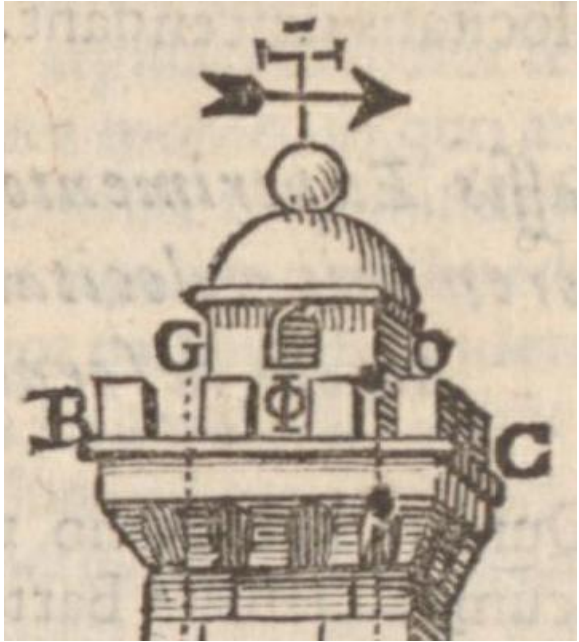

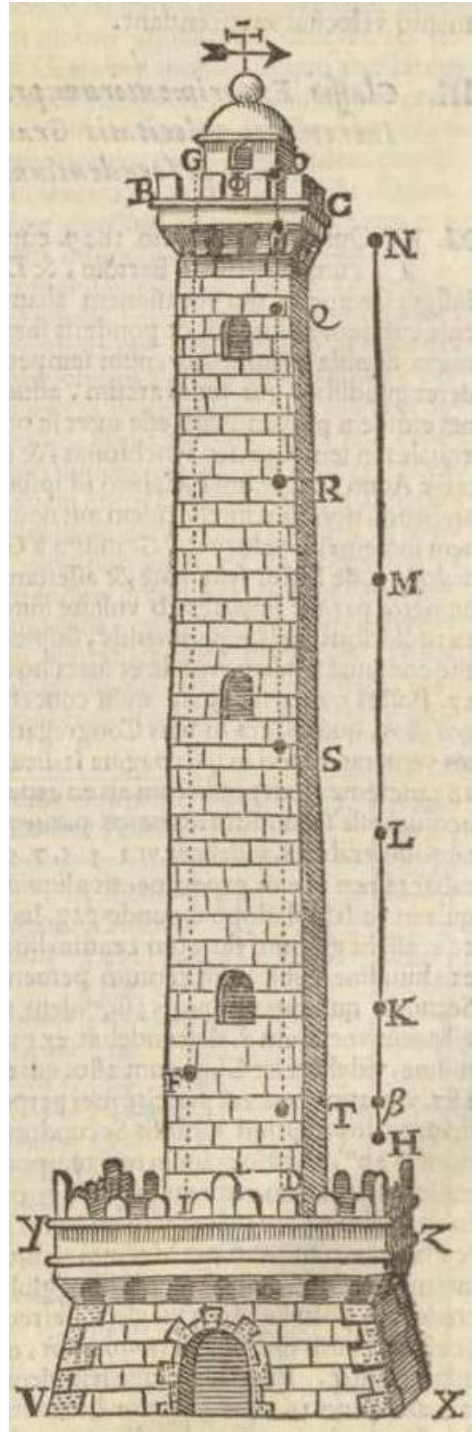

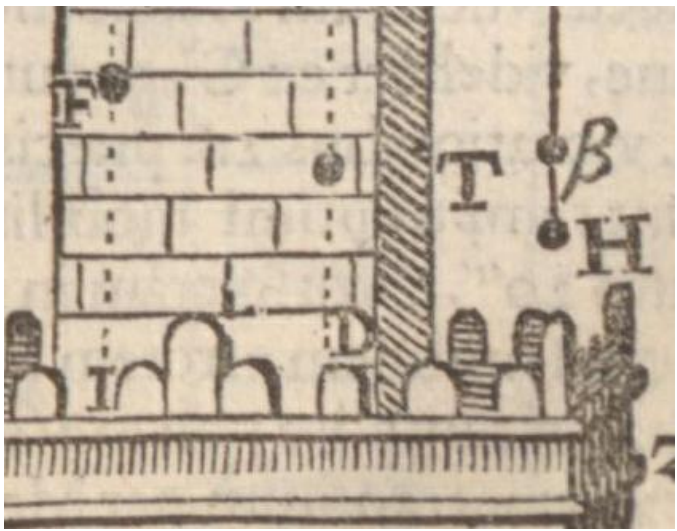

X. So in May of 1640, and at other times afterward, we determined the height Hβ, that from which an eight ounce clay ball, when released, will strike the pavement at precisely five exact strokes of the pendulum described above (that is, 5/6 of a second of time).  Through oft-repeated experiments we have discovered this to be 10 Roman feet.  Then we determined the height necessary for a ball of the same type & weight to descend in twice as much time, or ten strokes.  We discovered this to be 40 feet, which interval KH marks.  Ascending further, we determined the appropriate height for thrice as much time, or exactly 15 strokes, which we discovered to be 90 feet, LH.  We discovered that for a time of 20 strokes the height MH to be 160 feet, and for 25 strokes the height NH to be 250 feet.  Finally, we could not ascend sufficiently high for a ball to require 30 strokes for its descent.  So instead we repeatedly released a ball from the crown of the Asinelli tower at G to strike the pavement at I, which is a distance of 280 feet.  With me at I and Frs. Grimaldi and Cassiani at G, we consistently counted 26 strokes, as we discovered by comparison of our written notes.

> Riccioli appears to have determined the distance a ball drops during 5 pendulum strokes, and then tested whether the distances dropped in 10, 15, 20 strokes follows a progression of square numbers as described by Galileo in his *Dialogue*.  Thus of the several measurements given here, only the 5 stroke/10 foot drop, and the last, 280 foot drop (which does not fit in the progression) appear to be fully independent measurements.  Riccioli's further discussions (seen later on in this paper) also suggest this.

Now let us imagine the intervals marked on the line NH, translated to intervals on the line OT.  The distance the ball has travelled at the end of the first five strokes, OC, is 10 feet, and equals βH; the distance the ball has travelled at the end of the second five strokes, OQ, is 40 feet, and equals KH; at the end of the third five, OR, equal to LH, is 90 feet; at the end of the fourth set OS, equal to MH, 160 feet; at the end of the fifth set or 25 strokes the total OT equals the whole NH, 250 feet.  Based on the motion prior to T you have some indication of what occurs as the ball continues into the pavement.

Therefore the aforementioned ball descends faster and faster the farther it recedes from O and the nearer it approaches to D.  In terms of equal measures of time, during the first measure it traverses OC, 10 feet; during the second it traverses CQ, 30 feet; third, QR, 50 feet; fourth, RS, 70 feet; and fifth, ST, 90 feet.  These numbers sum to 250 feet.  Any such conspicuous and noted growth is deserving confirmation by further experiment.  Indeed we examine this alone: whether Heavy bodies falling naturally through the air in a straight line perpendicular to the horizontal, descend at a speed that is uniform, or increasing or decreasing by uniform differences.

*III. The Group of Experiments about the Proportion of the Growth of the Velocity of Heavy Objects descending through the Air.*

XI. I did not understand or recognize the proportion of the growth of the velocity of Heavy bodies related by Galileo in the Second Day of his Dialogue of the system of the World, and asserted by him to be following odd numbers begun from unity.  This is true even though I might have discovered it myself, beginning in at least the Year 1619 during an occasion when I was with Fr. Daniel

Bartolo, and Dr. Alphonso Iseo the eminent Geometer (examining two pendulums of the same height and weight simultaneously released from the same terminus, and whether they might always advance by like pace through any number of swings; noting how all oscillations of the same pendulum are mutually equal in time, or synchronous), and again later in 1634 with Fr. Cabeo at Ferrara. Indeed, at that time, according to my ruder experiments, I suspected it to be continually triple, as in these numbers: 1, 3, 9, 27.

Yet still later the opportunity was granted to me of reading Galileo's dialogues, which the Holy Congregation of the Index had prohibited. I found in the dialogues on page 217 of the Italian or 163 of the Latin[5] the aforementioned growth, discovered by experiment, to be following simple odd numbers from unity, as in 1, 3, 5, 7, 9, 11, etc. Still, I was suspecting something fallacious to lurk in the experiments of Galileo, because in the same dialogue, following page 219 of the Italian, 164 of the Latin[6] he asserts an iron ball of 100 Roman pounds released from an altitude of 100 cubits reaches the ground in 5 Seconds time. Yet the fact was that my clay ball of 8 ounces was descending from a much greater altitude, namely from GI (280 feet, which is 187 cubits) in precisely 26 strokes of my pendulum: 4 and 1/3 Seconds time. I was certain that no perceptible error existed in my counting of time, and certain that the error of Galileo resulted from times not well calibrated against transits of the Fixed stars — error which was then transferred to the intervals traversed in the descent of that ball. Furthermore, I was scarcely believing that Galileo had been able to use an iron ball of such great weight, especially when he

[5] See Galilei & Drake 2001, 257.

[6] See Galilei & Drake 2001, 259.

did not even name the tower from which he might have arranged for such a ball to be released.

And so, full of this suspicion, I began exacting measure of this growth in the Year 1640, as I have said. I hoped to contrive my own idea about this that was nearer to the truth; but rather I have in fact discovered to be true that which Galileo asserted. And indeed as I set forth in the preceding experiment (paragraph number X), I have acknowledged the growth to follow the proportion of 10, 30, 50, 70, 90 feet, which expressed in smallest numbers is just 1, 3, 5, 7, 9.

But not yet completely acquiescing to that, I examined with Father Grimaldi the height required in order that the eight ounce clay ball when released might reach the pavement in 6 strokes of the pendulum, or one Second; we obtained the height $\beta H$ to be 15 feet. For two Seconds, or 12 precise strokes, the height KH, was 60 feet. For three Seconds or 18 strokes the height LH was 135 feet. For 24 strokes or four Seconds the height MH was 240 feet. The same ball again and often transited the 280 feet at the Tower of Asinelli in 26 strokes or 4 and 1/3 Seconds.

I have discovered the preceding number of strokes to exactly correspond to the preceding distance intervals, although in the greater distances one or another foot less or more does not introduce a

> Here Riccioli again provides a simple estimate of experimental error.

difference of one whole stroke. Hence this experiment shows the proportion for distances traversed in equal times to have been 15, 45, 75, 105 feet, which follows 1, 3, 5, 7 (for as 1 is to 3, so is 15 to 45, etc). The results of these two experiments (and a third which I have not discussed here, both for the sake of brevity and because fraction numbers are involved) are found in the following table.

| Order of Experiments | Strokes of a Pendulum of length 1 & 15/100 inches | Time corresponding to the strokes (in seconds) | Square of the number of Strokes | Distance traversed by an 8-ounce clay ball at the end of the times (in Roman feet) | Distance traversed during each equal interval of time (in Roman feet) | Proportion of the growth of the velocity of heavy bodies in air (smallest numbers) |
|---|---|---|---|---|---|---|
| | 5 | 5/6 | 25 | 10 | 10 | 1 |
| 1 | 10 | 1 2/3 | 100 | 40 | 30 | 3 |
| | 15 | 2 1/2 | 225 | 90 | 50 | 5 |
| | 20 | 3 1/3 | 400 | 160 | 70 | 7 |
| | 25 | 4 1/6 | 625 | 250 | 90 | 9 |
| | 6 | 1 | 36 | 15 | 15 | 1 |
| 2 | 12 | 2 | 144 | 60 | 45 | 3 |
| | 18 | 3 | 324 | 135 | 75 | 5 |
| | 24 | 4 | 576 | 240 | 105 | 7 |
| | 26 | 4 1/3 | 676 | 280 | 40 | 8 1/6 |
| | 6 1/2 | 1 | 42 | 18 | 18 | 1 |
| 3 | 13 | 2 1/6 | 169 | 72 | 54 | 3 |
| | 19 1/2 | 3 1/4 | 380 | 162 | 90 | 5 |
| | 26 | 4 1/3 | 676 | 280 | 118 | 6 1/18 |

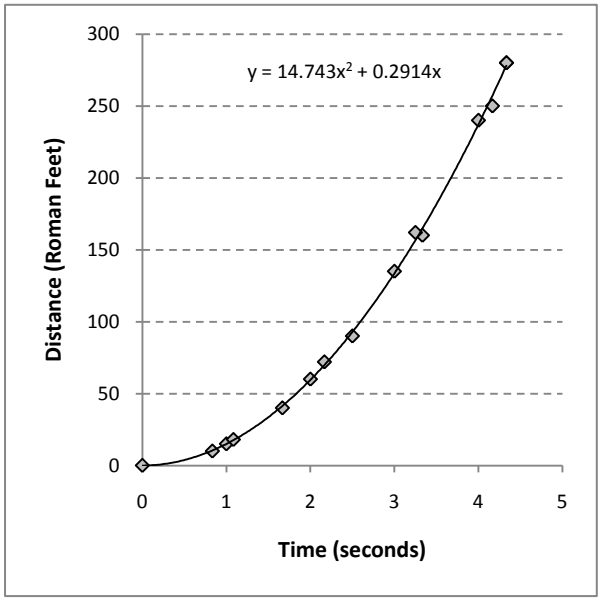

$y = 14.743x^2 + 0.2914x$

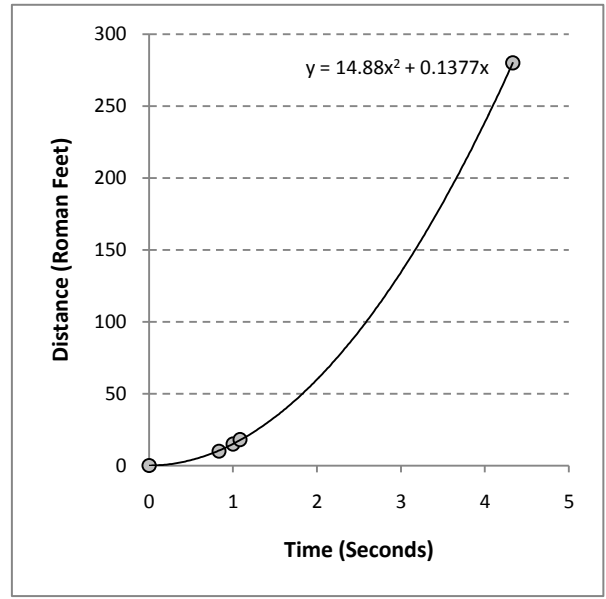

$y = 14.88x^2 + 0.1377x$

**Left:** Plot of Riccioli's data from his table above. The fit to the data yields acceleration due to gravity (g) of 29.5 Rmft/s$^2$. **Right:** Riccioli appears to have measured the time of fall for three different heights, and then tested for multiples of those times (and confirmed) Galileo's idea that distance increases as time squared; only the initial measurements, and the measurement from the top of the Asinelli Tower, are fully independent. These four independent measurements (highlighted in the table) are plotted here. The fit to these data yields g = 29.8 Rmft/s$^2$. Calculating g using d = ½ g t$^2$ and the four independent measurements yields g = 29.8 Rmft/s$^2$ with a standard deviation of 0.7 Rmft/s$^2$. If Riccioli's Roman foot is that measurement commonly given today as 0.296 m, then the accepted value of g would be 33.11 Rmft/s$^2$ (9.8 m/s$^2$) and Riccioli's value is off by about 9% from the accepted value. Thus for the 280 Rmft drop, Riccioli "should" have measured a time of 24.7 strokes rather than the 26 strokes he did measure, (continued next page…)

Therefore Fr. Grimaldi and I went to talk to the distinguished Professor of Mathematics at the Bologna University, Fr. Bonaventure Cavalieri, who was at one time a protégée of Galileo. I told him about the agreement of my experiments with the experiments of Galileo, at least as far as this proportion. Fr. Cavalieri was confined by arthritis and gout to a bed, or to a little chair; he was not able to take part in the experiments. However it was incredible how greatly he was exhilarated because of our testimony.

XII. Now those not ignorant of Geometry recognize the distances traversed by naturally descending heavy bodies of this sort increase as the squares of the elapsed times of descent. Galileo himself notes this in day 2 of the Dialogue page 217 or 163.[7]

For instance, in the first experiment, the strokes followed the sequence 5, 10, 15, 20, 25, the square numbers of which (that is, the products of the same number with itself) are 25, 100, 225, 400, 625; while the distances traversed followed the sequence 10, 40, 90, 160, 250. Now as 25 is to 100, so 10 is to 40; as 100 is to 225, so 40 is to 90; as 225 is to 400, so 90 is to 160; finally as 400 is to 625, so 160 is to 250. Thus in experiment 2, the order of Strokes was 6, 12, 18, 24, 26 the squares of which are 36, 144, 324, 576, 676. Truly the distances traversed were in sequence 15, 60, 135, 240, 280. But as 36 is to 144, so 15 is to 60; as 144 is to 324, so 60 is to 135; as 324 is Upon reading this paragraph the modern Reader will doubtlessly have a great appreciation for the ease, convenience, and brevity of today's mathematical notation!

---

[7] See Galilei & Drake 2001, 257.

to 576, so 135 is to 240; finally as 576 is to 676, so 240 is to 280. Therefore the preceding distances have increased in proportion to the squares of the strokes (that is, times). Reducing those times to least numbers, so that the first time is 1 unit, the second 2, the third 3, the fourth 4, and the fifth 5, the squares progress as 1, 4, 9, 16, 25.

In the third experiment (which you have available in the table seen here) the number of feet [on the last entry] ought to be exactly 288 in order that the preceding proportion might be preserved. However, the greatest height we were able to test was 280 feet.

It is truly pleasing to collect all that we have thus far said about so beautiful a proportion, and the basis of it, into one table that provides the Reader with a short glancing synopsis. Yet I add that at one time I was hoping we would discover this same proportion in the weight elevated by a falling ball [striking one side of a balance] — so that if a ball falling from a height of one foot raised a weight an inch, it would raise the weight doubly much from the height of 4 feet, or triply much from the height of 9 feet. It did not. See the upcoming group XII of the Experiments.

> **The amount that a falling ball that strikes one pan of a balance will elevate a weight in the other pan would correlate to what today would be considered the Energy of the ball. Riccioli had expected this value to proceed as the square of the height from which the ball falls. However, the gravitational potential energy (and thus the Kinetic Energy upon impact) of a falling object increases directly with height.**

In conclusion, Riccioli's experiment speaks for itself.  He is thorough.  He provides a full description of his experimental procedure.  His data — which includes estimates of uncertainty for his measurements — is of sufficient quality to determine the acceleration due to gravity (g) to an accuracy of 5%.  He undertakes the experiment expecting to disprove Galileo's ideas (distrusting Galileo in part because Galileo does not report the sorts of experimental details that he would); yet when his results instead confirm Galileo's ideas, Riccioli makes a point of promptly sharing the exciting news with an interested colleague who worked with Galileo, and later publishing the results.  Riccioli appears to be a model scientist.